\RequirePackage[l2tabu, orthodox]{nag}
\documentclass[twocolumn,amsmath,amssymb,aps,showkeys,nofootinbib]{revtex4-2}

\usepackage{graphicx}% Include figure files
\graphicspath{{figures/}}
\usepackage{dcolumn}% Align table columns on decimal point
\usepackage{bm}% bold math
\usepackage[hidelinks]{hyperref}% add hypertext capabilities
\usepackage{titlesec}
\titleformat{\section}[hang]{\normalfont\bfseries}{\thesection.}{0.5em}{}[]
\titlespacing{\section}{0pt}{*2}{*1}

\begin{document}

%% TITLE + AUTHOR STUFF
\title{A modeling approach for estimating dynamic measles case detection rates}
\author{Niket Thakkar}
%\email{niket.thakkar@gatesfoundation.org}
\homepage{https://nthakkar.github.io/}
\homepage[\\]{https://github.com/NThakkar-IDM/uk_measles_surveillance}
\affiliation{%
The Institute for Disease Modeling\\
at the Bill $\&$ Melinda Gates Foundation\\
Seattle, Washington 98122
}%
\date{\today}% It is always \today, today,
             %  but any date may be explicitly specified

%% ABSTRACT
\begin{abstract}
The main idea in this paper is that the age associated with reported measles cases can be used to estimate the number of undetected measles infections. Somewhat surprisingly, even with age only to the nearest year, estimates of underreporting can be generated at the much faster, 2 week time-scale associated with measles transmission. I describe this idea by focusing on the well-studied, 60 city United Kingdom data set, which covers the transition to universal healthcare in 1948, and is, as a result, an interesting case study in infectious disease surveillance. Finally, at the end of the paper, I comment briefly on how the approach can be modified for application to modern contexts.
\end{abstract}
%Use showkeys class option if keyword display desired
\maketitle

%% MAIN TEXT
\section{\label{sec:intro}Underreporting infections as a dynamic process in epidemiology}
When we try to make decisions to mitigate infectious disease burden, whether that's in allocating vaccines or in applying other interventions, the key piece of data informing our epidemiological perspective is often reported infections over time. In many situations, this timeseries can be stratified across social dimensions, like age, or location, or sex, and differences in trends are then used to make directional statements regarding risk (``cases are rising faster in men than in women", for example). These risk assessments ultimately become one part of a broader decision-making conversation.

It's almost always the case that a significant fraction of infections go undetected \cite{bedford2020cryptic}. This reality often forces epidemiologists to assume underreporting happens uniformly over time and strata, facilitating at least some \textit{relative} estimates of burden and risk \cite{goldstein2020temporal}. While this assumption might be good in certain circumstances, it typically goes untested, and its validity is challenged by other realities of healthcare, like inequities and disruptions in access across populations \cite{martinez2020sars,who2018}.

Estimating the degree of underreporting is a challenging problem. Surveillance data, despite being so central to our epidemiological understanding, usually comes from convenience samples and is rarely collected through formal, randomized surveys. Data quality is then, at best, estimated through independent serological surveys that measure population immunity, a point of comparison to the level of immunity implied by accumulated case reports and data on vaccination \cite{winter2018benefits}. This approach can give some assessment of underreporting across social dimensions, but since serosurveys can't be performed continuously, it has fundamental limitations when it comes to estimating changes over time.

Transmission modeling offers a complementary approach, but it comes with its own challenges. At a high level, transmission models consider interactions between a population's infectious and susceptible individuals over time \cite{anderson1992infectious}, offering a mathematical platform for interpreting surveillance data in the context of an underlying disease transmission process. In principle, this approach yields time-resolved estimates of the total infectious population which can be compared to case reports to dynamically assess underreporting. In practice, however, transmission models are under-constrained, and the same surveillance data can be explained by multiple, distinct combinations of reporting and transmission processes. 

Resolving this identifiability issue, specifically in the context of endemic diseases like measles, is the focus of this paper. Broadly speaking, I find that the year-to-year changes in the surveillance system can be constrained by the age distribution of reported cases and then used to regularize the behavior of a high time-resolution transmission model. This approach yields a model with sufficient structure to distinguish reporting and transmission volatility while avoiding assumptions about the fast-time-scale variation in reporting. As a result, it offers a general procedure for dynamic assessment of the surveillance system based on reported cases over time.

To be concrete, I describe the method by way of example. In particular, I focus on the well-studied, 60 city United Kingdom (UK) dataset, which contains biweekly reports of measles cases from 1944 to 1966, covering the establishment of the National Health Service (NHS) in 1948 but before the introduction of the measles vaccine in 1967. The lack of vaccine, of course, limits direct applicability of the results to modern contexts, but it helps to simplify the discussion and focus on disease surveillance, which changed significantly with the nationalization of healthcare. As a result, this example proves to be an illustrative case study of the interactions between a variety of social forces as viewed through disease transmission.

\section{Measles in the UK from 1944 to 1966}
%% FIGURE 1: DATA VIS
\begin{figure*}
\centering
\includegraphics[scale=0.56]{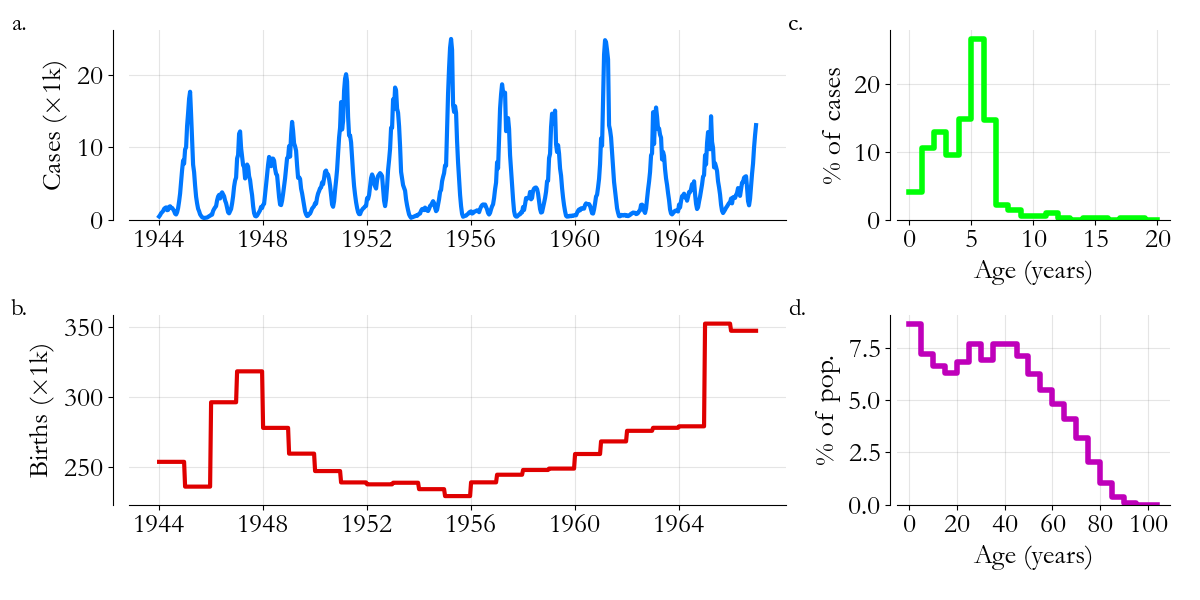}
\caption{\label{fig:data}Data from the UK (1944-66). \textbf{(a)} Biweekly measles case reports from 60 cities, \textbf{(b)} live births per year in the same cities, \textbf{(c)} the age distribution of measles cases in 1950, and \textbf{(d)} the age-pyramid of the population in 1950 are the main inputs used to construct a transmission and reporting model.}
\end{figure*}

The key pieces of data for this study are visualized in Fig. \ref{fig:data}. In panel a, biweekly measles case reports were taken from Registrar General's \textit{Weekly Reports} by the Grenfell group and made open access. Similarly, Grenfell's group compiled yearly live-birth estimates, shown in panel b, from the Registrar General's \textit{Annual Reports} \cite{finkenstadt2000time}. For the purposes of this paper, I added to this dataset Fine and Clarkson's distribution of the ages of measles cases circa 1950 (Fig. \ref{fig:data}c) \cite{fine1982measles2} and an estimate of the population pyramid for the same year (Fig. \ref{fig:data}d) \cite{poppyramid}. 

In many ways, this is \textit{the classic} measles dataset, and a lot of its features are very well understood. For example, transmission modeling has established a dynamic connection between the baby-boom after World War 2 (1946-48) and the shift from biannual to annual measles outbreaks from 1947 to 1953 \cite{grenfell2002dynamics}. Similar studies have also established a connection between seasonal variation in transmission rates and the holiday calendar for UK schools \cite{finkenstadt2000time,fine1982measles1}. Disaggregating the data across the 60 cities has facilitated understanding of the population size needed to maintain endemic transmission \cite{keeling1997disease} and the wave-behavior associated with measles' diffusion from city-to-city \cite{grenfell2001travelling}. In short, many researchers have brought a variety of perspectives to this data, teaching us a lot about measles epidemiology in the process.

That said, to my knowledge, no study has focused on the time-variation in reporting rates associated with this data.\footnote{Researchers have made estimates, for example in Ref. \onlinecite{finkenstadt2000time}. But in that paper, the estimates are unpublished and not discussed.} This is surprising since, with the establishment of the NHS in July 1948, the population's relationship to healthcare changed dramatically \cite{nhshistory}. Given the level of societal detail apparently reflected in measles' transmission dynamics, it's reasonable to expect that so profound a change should be observable as well. 

\section{Modeling measles dynamics\label{sec:problem}}

This thought can be framed with more mathematical precision by defining a discrete stochastic process model of measles transmission and reporting. Along the lines of classic disease models \cite{anderson1992infectious,finkenstadt2000time}, and mindful that measles is transmitted person-to-person, the time from exposure to rash onset is roughly 14 days, and those that survive the disease are immune for life, I assume
\begin{align}
    I_t &= \beta_{t-1}S_{t-1}I_{t-1}^{\alpha}\varepsilon_{t-1},\label{eq:transmission}\\
    S_t &= S_{t-1}+B_{t-1}-I_{t},\label{eq:susceptibles}\\
    C_t &\sim \text{Binomial}\left\{I_t,r_t\right\} \label{eq:reporting}
\end{align}
where, at time $t$ in 2 week time steps, $S_t$ is the population susceptible to measles, $I_t$ is the infectious population, $B_t$ are births into the population (e.g., Fig. \ref{fig:data}b, interpolated to the biweekly time scale), and $C_t$ are reported measles cases (e.g., Fig. \ref{fig:data}a).

New infections are generated through a transmission process where $\beta_t$ is the average transmission rate and $\varepsilon_t$ models multiplicative transmission volatility (i.e., $\text{E}[\ln\varepsilon_t]=0, \text{V}[\ln\varepsilon_t]=\sigma^2_t$). These quantities, taken together, select a random fraction of possible susceptible-infectious pairs to contribute to onward transmission and replace the previous infectious generation. Notice, however, that $I_{t}$ can be discounted by $\alpha\leq1$ in the enumeration of pairs, modeling the idea that as $I_t$ grows, the likelihood of infection among relatively isolated children, who are incapable of onward transmission, like the very young, increases.

The reporting process, $r_t$, represents the probability that infections are reported as cases and is in principle free to vary such that $0\leq r_t \leq 1$. Evaluating a measles surveillance system in the context of this model is therefore a statistical inference problem where the goal is to estimate unknowns $r_t$, $\beta_t$, $\varepsilon_t$, $\alpha$, and $S_0$ given $C_t$ and $B_t$. 

As mentioned, this inference problem is poorly-posed in general. To illustrate the issue, consider solving the expected value of Eq. \ref{eq:reporting} for $I_t$ and inserting the result into Eq. \ref{eq:transmission} to write
\begin{align*}
    \frac{C_t}{C_{t-1}} = \left(\frac{r_t}{r_{t-1}}\right)\beta_{t-1}S_{t-1}\varepsilon_{t-1},
\end{align*}
where I've set $\alpha=1$ for clarity. If, for example, $C_t$ increases relative to $C_{t-1}$ so that $C_t/C_{t-1} > 1$, that variation can either be explained by sufficient increase in $r_t$ or by a sufficiently high reproductive number, $\beta_{t-1}S_{t-1}\varepsilon_{t-1}$. In other words, in the context of an imperfectly observed epidemiology, an increase in reported cases can either be a result of more people seeking care by chance or an unfortunately social infectious generation or both. Without additional model structure and information, we cannot tell these effects apart. 

\section{Constraining systemic changes\label{sec:prior_construction}}

Some qualitative considerations can help motivate a way forward. In general, we expect $r_t$ to have 2 distinct timescales: A fast timescale, comparable to the 14 day transmission scale, associated with random fluctuations in the health-seeking behavior of $I_t$, and a much slower timescale associated with changes in the health system like the creation of new facilities. While it's difficult to say in advance how ``slow" should be defined, we also expect $\beta_t$ to vary seasonally with environmental conditions, and as a result, at any sub-annual timescale, we should anticipate seeing both transmission and reporting variation overlaid. The year-to-year timescale is perhaps the highest resolution where we can reasonably expect systemic reporting variation to be isolated.

The age-at-infection distribution in Fig. \ref{fig:data}c characterizes the UK's measles transmission at the annual scale. The two peaked structure is consistent with school-driven transmission, the larger peak corresponding to infection at school entry at roughly 5 years old and the smaller peak likely including the 5 year-olds' younger siblings at home. While the distribution is based on reported cases in 1950, this mechanistic connection to the school system gives some confidence that it can be taken as representative of the infectious population as a whole for the entire 23 year period of interest.\footnote{Arguably, this type of long-term stability in the age-distribution is the quantitative definition of endemic. In any case, I only have the one age distribution from 1950, and so I make the best of it.}

%% FIGURE 2: coarse inference
\begin{figure*}
\centering
\includegraphics[scale=0.54]{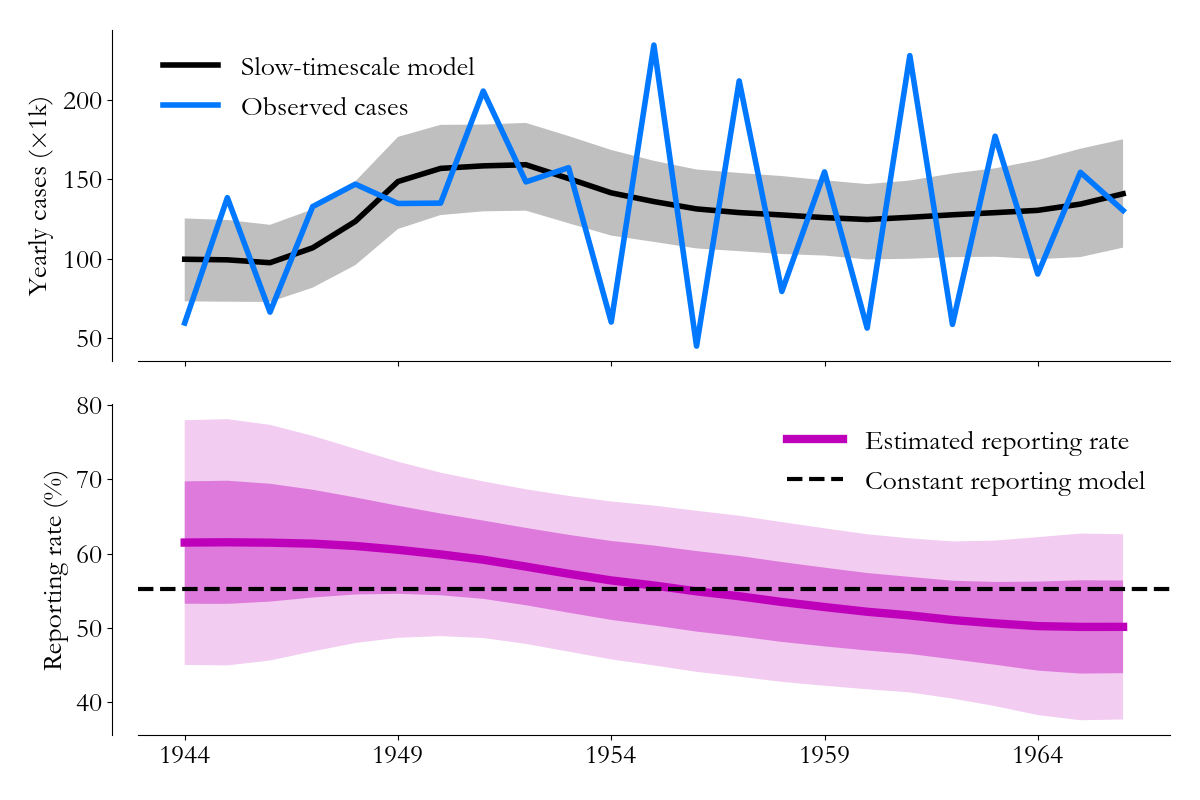}
\caption{\label{fig:slowmodel}Slow, systemic changes in surveillance. (Top) Age-at-infection information yields a model (black, 95$\%$ interval in grey) of total burden that can be compared to aggregated cases (blue) to estimate $\tilde{r}_t$. (Bottom) Visualizing the resulting estimate (purple, $50\%$ interval light, $95\%$ interval lightest) exposes an overall downward trend from 1944 to 1966.}
\end{figure*}

More formally, I'll assume the distribution in Fig. \ref{fig:data}c, $\pi(a)$, is the probability of rash onset at age $a$, or equivalently, the probability of infection, $i(t|s)$, during year $t=a+s$ given birth year $s$. Marking annually aggregated quantities with a tilde, I can compute
\begin{align}
    E[\tilde{I}_t^B] &= \sum_{s=0}^{T-1} i(t|s)\tilde{B}_s=\sum_{s=0}^{T-1}\pi(t-s)\tilde{B}_s,\label{eq:newborn_inf}
\end{align}
where $\tilde{I}_t^B$ are infections in children born during the $T=23$ years of interest. This interpretation of the age-at-infection distribution, that it tells us how birth-cohorts are expected to appear as infections over time, is the key idea of this paper. It inspires a strategy to compute total expected burden, $\text{E}[\tilde{I}_t]$, as a point of comparison to $\tilde{C}_t$.

Eq. \ref{eq:newborn_inf} accounts for one of two possible sources of infections in our model. If susceptible individuals were not born in the years 0 to $T-1$, they must have been part of $S_0$, the initial susceptible population. In other words, calculating expected infections in $S_0$, $E[\tilde{I}_t^{0}]$, would complete the estimate of total burden over time.

% THIS MIGHT BE SLIGHTLY WRONG??
The age-at-infection distribution offers valuable perspective again. If we approximate the transmission process as a simple random sample of the susceptible population, sometimes called a ``well-mixed" assumption, and we further assume that $\pi(a)$ is representative of years before year 0, then the initially susceptible population has age distribution $p(a|\in S_0) =\pi(a)$ as well. 

Operating under this approximation, inspired by the logic of Eq. \ref{eq:newborn_inf}, $S_0$ can be thought of as a mixture of birth-cohorts born in years $s<0$. This implies that
\begin{align*}
    \text{E}[\tilde{I}_t^0] &= S_0 \sum_{s<0} i(t|s)p(-s|\in S_0)\\
    &= S_0\sum_{a>0}\pi(t+a)\pi(a) \equiv S_0\tau(t),
\end{align*}
where $\tau(t)$ is the convolution of $S_0$'s age-distribution with the age-at-infection distribution --- the discrete self-convolution of $\pi(a)$ under our approximations. Moreover, the number $S_0$ can be constrained by the survival function, $\Pi(a) = 1 - \sum \pi(a')$, estimating the probability of remaining susceptible at age $a$. Combined with the age pyramid, $n(a)$, in Fig. \ref{fig:data}d, this means that we expect $S_0$ to be mostly individuals less than 5 years old, roughly $7\%$ of the total population.

Taken together then, expected yearly burden, $E[\tilde{I}_t] = E[\tilde{I}_t^B + \tilde{I}_t^0]$, can be estimated entirely from yearly births, the age-at-infection distribution, and the age-pyramid. The annualized reporting rate then satisfies 
\begin{align}
    \tilde{C}_t = \tilde{r}_t\left(\text{E}[S_0]\tau(t) + \sum_{s=0}^{T-1} \pi(t-s)\tilde{B}_s\right) + \tilde{w}_t,\label{eq:slowmodel}
\end{align}
with additive noise $\tilde{w}_t$. We can enforce the constraint $0 \leq \tilde{r}_t \leq 1$ by modeling $\tilde{r}_t = f(\tilde{\theta}_t)$ where $f(\cdot)$ is the logistic function and $\tilde{\theta}_t$ is a Gaussian process (see Appendix A). Then, modeling the variance in $\tilde{w}_t$ as a constant fully specifies a non-linear least squares problem that can be solved for $\tilde{\theta}_t$ to estimate $\tilde{r}_t$. 

Fig. \ref{fig:slowmodel} visualizes the results of this approach applied to the data from the UK. In the top panel, $\tilde{r}_t\text{E}[\tilde{I}_t]$ (black) follows the trend in $\tilde{C}_t$ (blue) with uncertainty ($95\%$ interval in grey) driven largely by the biannual periodicity in outbreaks starting in 1954. Eq. \ref{eq:slowmodel} is clearly an incomplete epidemiological model, lacking the transmission process required to explain outbreaks, and somewhat reassuringly, it cannot capture key features of the data with reporting variation alone.

But still, this is progress. The corresponding distribution for $\tilde{r}_t$ is visualized in the lower panel, with the $50\%$ and $95\%$ intervals in progressively lighter tints. The estimate is consistent with a constant-reporting-rate model (black) \cite{finkenstadt2000time}, but captures an overall falling trend from 1944 to 1966. More practically, the range of probable $\tilde{r}_t$ values, which was initially only loosely constrained by the rules of probability, is dramatically reduced through implications of the age-at-infection distribution.

\section{Balancing slow and fast dynamics\label{sec:inference}}
Keeping these results in mind, we can return to the more general inference problem outlined in Section \ref{sec:problem}. In somewhat abstract terms, completely defining Eqs. \ref{eq:transmission} to \ref{eq:reporting} requires us to calculate the posterior probability distribution $p(\beta_t,\varepsilon_t,\alpha,r_t,S_0|\mathbf{D})$, where $\mathbf{D}$ is the complete dataset $\{C_t,B_t,\tilde{C}_t,\tilde{B}_t,\pi(a),n(a)\}$, making clear the assumed separation of time scales. This formal statement of the problem is useful because it can be organized hierarchically to inspire an approachable inference algorithm. Specifically, we can write
\begin{align}
\begin{split}
    p(\beta_t,\varepsilon_t,\alpha&,r_t,S_0|\mathbf{D})=\\
    &p(S_0,r_t|\mathbf{D})\times p(\beta_t,\varepsilon_t,\alpha|\mathbf{D},S_0,r_t),\label{eq:posterior}
\end{split}
\end{align}
choosing this (exact but non-unique) separation to draw distinction between the parameters explicitly connected to the annual-scale ($S_0$, $r_t$) and the parameters responsible for fast dynamics $(\beta_t,\varepsilon_t,\alpha)$. As we'll see, the two terms on the right-hand-side lend themselves to approximation more readily than the distribution as a whole.
%because I want to leverage the constraints on $r_t$ and $S_0$ developed in the previous section. As we'll see, the two terms on the right-hand-side lend themselves to approximation more readily than the distribution as a whole.

The first term can be used as a vehicle for the survival analysis and annual-scale regression underlying the results in Fig. \ref{fig:slowmodel}. Towards that end, we can approximate
\begin{align*}
   p(S_0,r_t|\mathbf{D})&\approx p\left(S_0|\pi(a),n(a)\right)p(r_t|\pi(a),\tilde{C}_t,\tilde{B}_t,S_0)\\
   &\approx\mathcal{N}\left(S_0|\text{E}[S_0],\text{V}[S_0]\right)\mathcal{N}\left(r_t|\text{E}[\tilde{r}_t],\text{V}[\tilde{r}_t]\right).
\end{align*}
Here, the first line is the conditional independence assumption that $S_0$ and $r_t$, in the absence of a transmission process, are determined by the annual-scale subset of $\mathbf{D}$. Then, using $\mathcal{N}$ to represent the normal distribution, Gaussian approximations to these two terms are exactly what was estimated in the previous section if we linearly interpolate $\text{E}[\tilde{r}_t]$ and $\text{V}[\tilde{r}_t]$ to the 2 week timescale.

For the final term in Eq. \ref{eq:posterior}, it's helpful to return to the relationships in the model. Notice first that Eq. \ref{eq:susceptibles} can be solved for $S_t$ to give
\begin{align*}
    S_t = S_0 + \sum_{i=0}^{t-1}(B_i - I_{i+1}),
\end{align*}
the intuitive result that the susceptible population is the total balance of newborns and exposures every time step. Furthermore, taking the log of Eq. \ref{eq:transmission} implies
\begin{align}
   \ln I_t - \ln S_{t-1} = \alpha\ln I_{t-1} + \ln\beta_{t-1} + \ln\varepsilon_{t-1}.\label{eq:transmission_regr}
\end{align}
And finally, through properties of the binomial distribution (see Appendix B), Eq. \ref{eq:reporting} implies that $E[I_t|C_t,r_t] =[(C_t +1)/r_t]-1$. Thus, conditional on $r_t$ and $S_0$, Eq. \ref{eq:transmission_regr} is very nearly a well-defined linear regression for $\alpha$, $\beta_t$, and $\varepsilon_t$, but with $26T$ equations and up to $(52T-1)$ unknowns --- too many for a unique solution. 

We can use measles' transmission seasonality as an epidemiologically reasonable way to reduce dimensionality. Along those lines, I'll assume that $\beta_t$ is a periodic Gaussian process with 1 year (26 time step) periodicity (see Appendix A for details). Furthermore, I'll assume that $\ln\varepsilon_t$ has constant variance, $\sigma^2_{\varepsilon}$. Those assumptions dramatically reduce the number of unknowns, to 28 in total, making Eq. \ref{eq:transmission_regr} a solvable linear regression. 

Returning to Eq. \ref{eq:posterior}, the approximations just discussed give us a route to evaluate the entire right-hand-side, and as a result, with enough patience, we could calculate whatever properties of that distribution that we might be interested in. That said, since the inference problem remains high dimensional ($26T+29$ unknowns), it's reasonable to look for a more digestible distribution. A standard approach is to construct a Gaussian approximation at the distribution's mode \cite{sivia2006data}.

That procedure is facilitated by the negative log posterior, $\mathcal{L}$, which can be minimized to find the mode and twice-differentiated at the minimum to estimate a covariance matrix. For our purposes, up to a constant,
\begin{align}
\begin{split}
    \mathcal{L}(\beta_t,\varepsilon_t,\alpha&,r_t,S_0) = \frac{26T-1}{2}\ln\hat{\sigma}_{\varepsilon}^2\\
    &+\frac{(S_0-\text{E}[S_0])^2}{2\text{V}[S_0]}+\sum_t \frac{(r_t-\text{E}[\tilde{r}_t])^2}{2\text{V}[\tilde{r}_t]}\label{eq:neglogl}
\end{split}
\end{align}
where $\hat{\sigma}_{\varepsilon}$ is the maximum-likelihood estimate of $\sigma_{\varepsilon}$ associated with the least-squares solution to Eq. \ref{eq:transmission_regr}. This equation nicely illustrates the balance we've achieved, with the first term corresponding to the model's ability to capture the fast dynamics and the next two terms enforcing consistency with the slow time scale.

Eq. \ref{eq:neglogl} also encapsulates a well-defined statistical inference algorithm. We first minimize $\mathcal{L}$ to construct a Gaussian approximation of Eq. \ref{eq:posterior}, and then properties of Gaussians can be leveraged to calculate marginal distributions or draw sample parameter sets consistent with the data.\footnote{Some practical details: I minimize $\mathcal{L}$ using the \texttt{scipy}'s implementation of the Broyden–Fletcher–Goldfarb–Shanno algorithm, with the gradient of $\mathcal{L}$ computed exactly instead of with the default finite-differences. Moreover, noting that $S_0$ being orders of magnitude larger than $r_t$ can cause stability problems, I solve instead for $\ln S_0$, modifying the associated term in Eq. \ref{eq:neglogl}. Finally, for efficiency reasons, I first solve the problem with $\alpha=1$ to find $S_0$ and $r_t$ and then solve Eq. \ref{eq:transmission_regr} more generally after the fact. In practice, this means only solving Eq. \ref{eq:transmission_regr} twice, instead of once per evaluation of $\mathcal{L}$.} Finally, for any candidate set of parameters, Eqs. \ref{eq:transmission} to \ref{eq:reporting} allow us to produce time series of $I_t$ and $S_t$ and then estimate quantities of epidemiological interest.

%% FIGURE 3: model fit
\begin{figure*}
\centering
\includegraphics[scale=0.63]{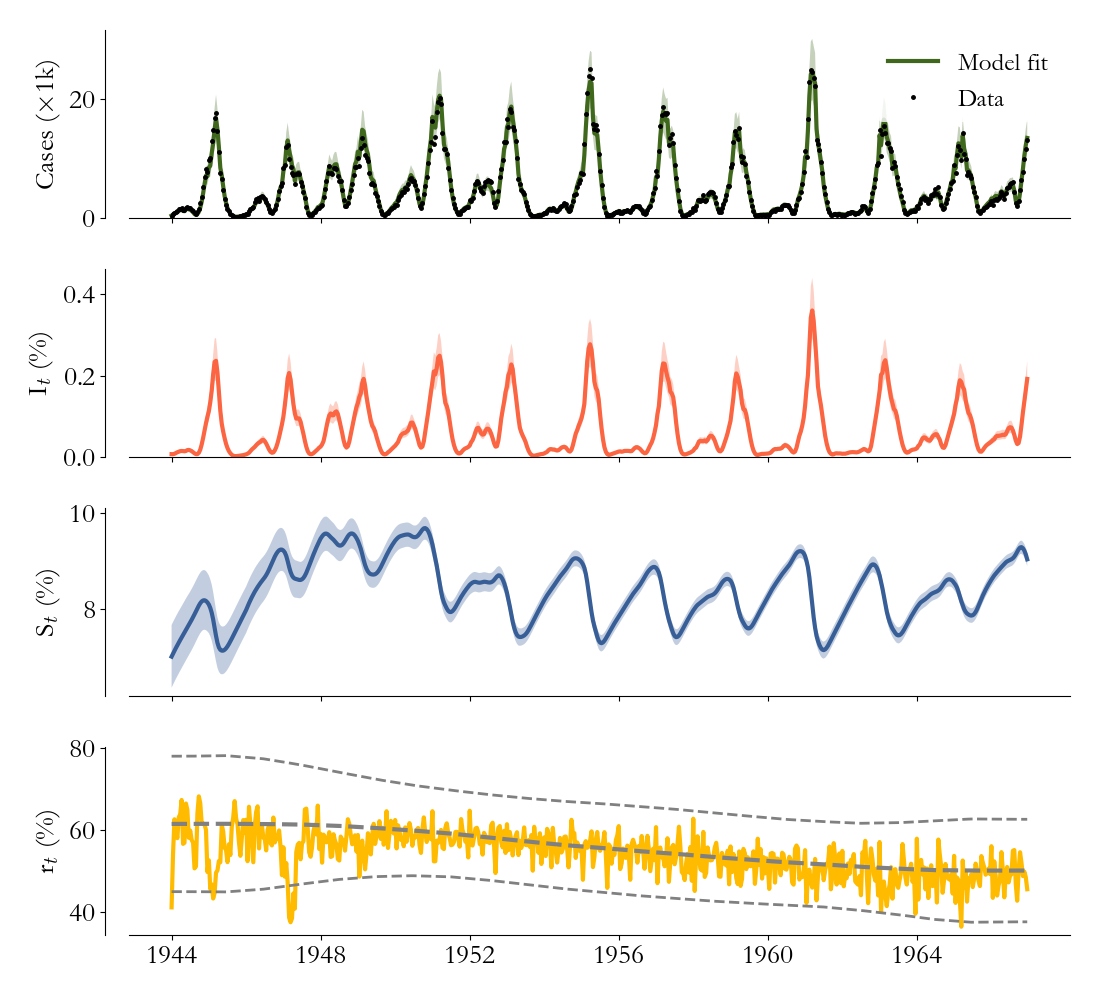}
\caption{\label{fig:fit}Modeling measles transmission and surveillance dynamics. Forcing consistency with reported cases, both in terms of the biweekly dynamics (black dots) and age-distribution-based results in Fig. \ref{fig:slowmodel} (dashed grey), yields a model (colors, 95$\%$ interval shaded) that can distinguish variation in population prevalence (peach) and susceptibility (blue) from the probability infections are reported (yellow).}
\end{figure*}

I apply this algorithm to the UK example in Fig. \ref{fig:fit}, using 10,000 sample trajectories drawn from the fitted model to quantify overall uncertainty. In the top panel, the model (green) follows $C_t$ (black dots) closely, now able to explain outbreaks with a seasonal measles transmission process. Moreover, the uncertainty estimates have good empirical performance, with the $50\%$ interval across trajectories capturing $53\%$ of the data and the $95\%$ interval capturing $96\%$ of the data.

We can also see good consistency with the results in Fig. \ref{fig:slowmodel}. In particular, the initial susceptible population (blue) is roughly $7\%$ of the total population, as expected based on the survival function. Even more clearly, in Fig. \ref{fig:fit}'s final panel, $r_t$ exhibits 2 distinct timescales, with step-to-step volatility associated with people's behavior and a year-to-year trend consistent with $\tilde{r}_t$ (overlaid in grey). Thus, speaking broadly, Fig. \ref{fig:fit} demonstrates that we've created a measles transmission model consistent with both the dynamics and the age distribution of reported cases, and in doing so, we've uncovered the variation in the surveillance system.

\section{Social forces as viewed through transmission}

\begin{figure*}
\centering
\includegraphics[scale=0.58]{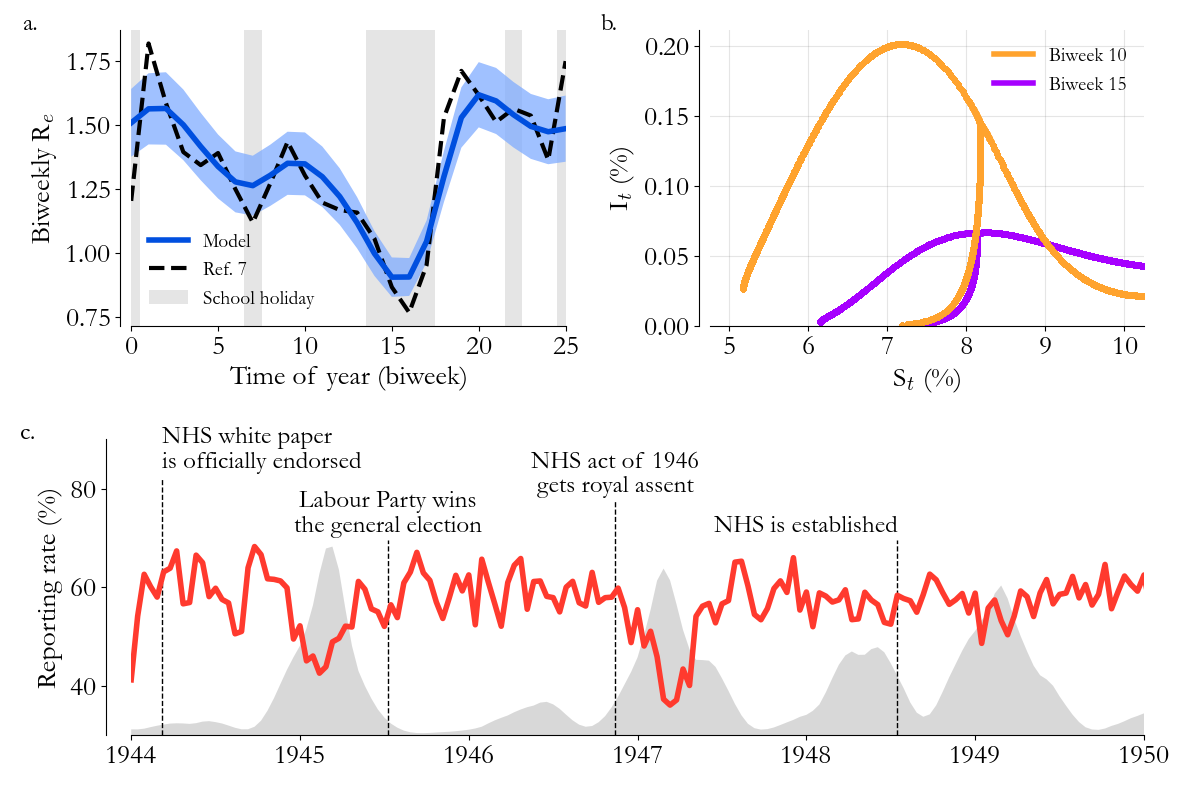}
\caption{\label{fig:results}Inferences from the fitted model. \textbf{(a)} The estimated seasonality profile (blue, 95$\%$ interval shaded) is correlated with the UK's school holiday schedule (grey), in good agreement with Ref. \onlinecite{finkenstadt2000time}. \textbf{(b)} Equilibrium analysis of trajectories at specific times of year (biweek 10 in orange, 15 in purple) can be used to construct a phase plane illustrating the deterministic, critical phenomena associated with the UK's outbreak periodicity. \textbf{(c)} Estimated $r_t$ (red) as healthcare is reformed, with time contextualized by the model's mean $I_t$ estimate (grey).}
\end{figure*}

Before discussing the $r_t$ estimates, it's worth establishing Fig. \ref{fig:fit}'s consistency with past studies. Two key results to that effect are visualized in Fig. \ref{fig:results}.

%% FIGURE 4: model inferences and healthcare access

In Fig. \ref{fig:results}a, I've plotted the model's estimate of the reproductive number (blue), which is proportional to $\beta_t$, in comparison with Ref. \onlinecite{finkenstadt2000time}'s point estimate (black) and the UK school holiday calendar (grey). The correlation with the school calendar is visually apparent, with closures suppressing transmission rates and the return from holidays associated with transient increases. Our estimate is noticeably smoother than Ref. \onlinecite{finkenstadt2000time}'s: This is because of the periodic Gaussian process mentioned in the previous section, modeling the fact that biweekly intervals occur at slightly different times of year every year. But outside of that minor difference, our estimates clearly recapitulate this classic result.

A second important result is the relationship between the post World War 2 baby boom (1946-48, see Fig. \ref{fig:data}b) and the transition from biannual to annual outbreaks. Following the approaches in Refs. \onlinecite{finkenstadt2000time} and \onlinecite{grenfell2002dynamics}, we can verify that this phenomena is a deterministic, dynamical feature of the fitted model by estimating the equilibrium periodicity of $I_t$ as a function of $S_t$ in expectation. 

More concretely, I can compute long time, 200 year trajectories of the model mean as a function of a constant birth rate, using the first century to reach equilibrium and the second century to sample it. Then, by plotting realized ($S_t$, $I_t$) pairs at specific times of year, I can construct phase portraits with geometric features that give insight into the model's dynamical properties.

The results of this procedure are visualized in Fig. \ref{fig:results}b. Choosing two representative times of year (biweek 10, in orange, in keeping with Ref. \onlinecite{finkenstadt2000time}'s choice, and biweek 15, in purple, because it's the peak of the low season), ($S_t$, $I_t$) pairs fall into 2 distinct phases. At low and high susceptibility, $I_t$ takes a single value at the same time every year, indicating that all years are the same and the equilibrium has annual periodicity. Meanwhile, in the vicinity of $8\%$ susceptibility, $I_t$ falls on one of two branches corresponding to low and high years -- that is, biannual periodicity.

Returning to the timeseries of Fig. \ref{fig:fit}, we see that $8\%$ susceptibility is indeed a critical, historically relevant threshold. In the post World War 2 years, increased birth rates pushed susceptibility to roughly $9\%$ and $I_t$ exhibits annual periodicity. At other times from 1944 to 1966, susceptibility hovered near $8\%$, and outbreaks occurred every other year. Fig. \ref{fig:results}b suggests that this transition was not due to chance, that is volatility in $\varepsilon_t$. Instead, it is a deterministic, in some sense physical, feature of the UK's measles epidemiology, reproducing the insight first explored in Ref. \onlinecite{finkenstadt2000time}.

The model's incorporation of the age-at-infection distribution adds some texture to these famous results. As mentioned, the distribution's two-peaked structure (Fig. \ref{fig:data}c) also supports the school system's role in transmission, suggesting that the correlation in Fig. \ref{fig:results}a can be interpreted causally. Meanwhile, that the 1946 to 1948 baby-boom raises susceptibility through 1952, that is for roughly 5 years, is what we might have expected based on the survival function, which vanishes after roughly 5 years. These elements of our inferences suggest that, even without directly modeling ageing, we've captured the dynamic implications of the age distribution.

Now, with some confidence in the corresponding transmission process, we can return to $r_t$. Fig. \ref{fig:results}c visualizes the estimate (red) in the lead up and aftermath of the NHS's establishment with the model's mean $I_t$ estimate (grey) to contextualize the time of year. It's striking that 1948 emerges naturally as a critical year in measles surveillance. Major outbreaks in 1945 and 1947 are accompanied by suppression in $r_t$, but this feature vanishes after 1948, and looking to Fig. \ref{fig:fit}, it does not return for the remaining 19 years. 

A plausible hypothesis explaining the full 23 year estimate is that, up to 1947, large-scale outbreaks placed significant burden on populations without access to healthcare. Then, after 1948, universal healthcare resolved this problem; however, the steady year-to-year $r_t$ decrease from 1948 onward suggests that increased demand impacted quality overall. Indeed, in the early 1970's, just after the model period, the NHS received it's first round of reforms in response to surprising demand \cite{nhshistory}. 

By virtue of being difficult to estimate, estimates of $r_t$ are difficult to validate. That said, the hypothesis above motivates a coarse prediction. 

If we assume that measles reporting is a product of two decisions, that an individual is sick enough to seek care and that they have access to care given their need, and we assign probabilities $p_N$ and $p_{A|N}$ to those two events, then post-1948, to a good approximation, $p_{A|N} = 1$. Meanwhile, we can further assume that $p_N$ is the same pre- and post-1948, since we expect it to be dominated by biological features of measles infections. As a result, we should expect the ratio of pre- and post-1948 $r_t$ estimates to measure $p_{A|N}$ before healthcare reform. 

And so, with roughly $60\%$ reporting after the NHS's establishment and roughly $40\%$ at peak suppression in 1947, we estimate that roughly $30\%$ of the UK's population lacked access to healthcare before reform. Looking to other historical records, this estimate is in good agreement with shortages of tuberculosis beds in 1947 (32,600 available with roughly 46,000 needed) \cite{nhshistory}, and survey data taken at that time \cite{nuffield1946hospital} might give additional validation -- I'm currently looking into it.

\section{Some final thoughts}

Zooming out to conclude: It's remarkable how disease transmission reflects society's layers with such clarity, from the school system to birth rates and finally to healthcare access. We've certainly seen this recently as well \cite{bedford2020cryptic}, sometimes harshly \cite{martinez2020sars}. But for our goals in measles control today, we should aspire to the level of detail with which we understand transmission and surveillance in the pre-vaccine-era UK.

Towards that end, this paper's approach can be extended to incorporate vaccination. At a high level, the annual burden estimates of Sec. \ref{sec:prior_construction} need to be adjusted for fractions of birth cohorts and of the initial susceptible population that are immunized before infection. Those adjustments can be informed by survey data on vaccination coverage, accounting for the administration age and, via the survival function, the probability of remaining susceptible at that age. Subtracting these estimates from $\text{E}[\tilde{I}_t]$ in Eq. \ref{eq:slowmodel} then allows $\tilde{r}_t$ to be constrained in essentially the same way. Inference at the fast timescale then proceeds as in Sec. \ref{sec:inference}, with Eq. \ref{eq:susceptibles} modified to account for immunization \cite{thakkar2019decreasing}. 

More generally, this paper has emphasized the need to understand not only the drivers of disease transmission but also the components of disease surveillance, thinking of both as dynamic, epidemiological processes. I hope retrospective, quantitative inference is a step towards understanding how surveillance systems can be improved.

\section*{Acknowledgements}
This work was done in conversation with many of my colleagues. I'd like to thank Safi Karmy-Jones for motivating the historical exploration, Edward Wenger for his intuition on the age-distribution's relationship to schools, and Mike Famulare for his sense of direction on the survival analysis. Edward, as well as Kevin McCarthy and Arie Voorman, also gave very useful feedback on this paper's first draft.

\appendix
%\vfill
\section{Gaussian processes}
The regression problems in Eqs. \ref{eq:slowmodel} and \ref{eq:transmission_regr} rely on Gaussian processes to create smoothness in time. More precisely, when I write that $\theta_t$ is a Gaussian process, I mean that $p(\theta_t) = \mathcal{N}(\theta_t|0,(\lambda \mathbf{D}^{\intercal}\mathbf{D})^{-1})$, where the matrix $\mathbf{D}$ is the finite-difference approximation to the second derivative. For periodic processes, $\mathbf{D}$ has periodic boundary conditions. Otherwise, the boundaries are handled by switching from centered to forward or backward approximations. In either case, this Gaussian distribution is taken as a prior in the relevant regression problems, leading to a penalty proportional to $\theta_t$'s second derivative and, in that way, enforcing smoothness. 

The constant $\lambda$ can be thought of in terms of correlation time. Specifically, the total variation, $\nu=||\mathbf{D}\theta_t||^2$, is Gamma distributed with shape $T/2$ and scale $2/\lambda$, so $\text{E}[\nu]$ is inversely proportional to $\lambda$. Meanwhile, for a sine wave with period $\tau$, the total variation goes as $\tau^{-4}$. Thus, in specifying $\lambda$, I choose an expected timescale for $\theta_t$ by setting $\lambda \propto \tau^4$. For Eq. \ref{eq:slowmodel}, I choose $\tau=5$ years, and for Eq. \ref{eq:transmission_regr}, I choose $\tau=3$ biweeks. That said, in sensitivity testing, none of the results of this paper change significantly for reasonable choices for $\lambda$.
    
\section{The binomial reporting model}
In specifying the model, I leverage one small theorem related to the binomial distribution, also discussed in Ref. \onlinecite{thakkar2019decreasing}. Specifically, if
\begin{align*}
p\left(C_t|I_t, r_t\right) = \binom{I_t}{C_t} r_t^{C_t}(1-r_t)^{I_t - C_t},
\end{align*}
then I can use Bayes' theorem with a uniform prior enforcing $I_t \geq C_t$ to compute
\begin{align*}
p\left(I_t | C_t, r_t \right) = \binom{I_t}{C_t} r_t^{C_t+1}(1-r_t)^{I_t - C_t},
\end{align*}
which is a distribution over $I_t$, normalized by the additional factor of $r_t$. The associated moment generating function is
\begin{align*}
    \text{E}[e^{sI_t}|C_t,r_t] = \frac{r_t^{C_t+1}e^{sC_t}}{[1 - (1-r_t)e^{s}]^{C_t+1}},
\end{align*}
which implies (by Taylor expanding and picking off the coefficient linear in $s$) that
\begin{align*}
\text{E}\left[I_t | C_t, r_t \right] &= \frac{C_t + 1}{r_t} - 1.
\end{align*}
This result relates Eq. \ref{eq:transmission_regr} to the reported data, helping us along the way to a well-defined linear regression for the transmission process. Incidentally, the dependence on $C_t+1$ makes $\ln\text{E}[I_t | C_t, r_t]$ well-defined even at $C_t = 0$ for any $0<r_t<1$, so this paper's methods are somewhat naturally capable, at least in this respect, of handling the sparse data typical of modern high-burden settings.

%% Make the bibliography
%\newpage
%\bibliographystyle{apsrmp}
\bibliography{references}
\end{document}